\documentclass[floats,preprint,aps,prb,showpacs,epsfig]{revtex4}
\usepackage{graphicx}
\usepackage{dcolumn}
\usepackage{bm}
\begin{document}
\preprint{APS/PRB}
\title
{Robust half-metallic antiferromagnets La$A$VOsO$_6$ and La$A$Mo$Y$O$_6$ 
($A$ = Ca, Sr, Ba; $Y$ = Re, Tc) from first-principles calculations}
\author{Y. K. Wang}
\affiliation{Center for General Education, Tajen Institute of Technology, 
Pingtung 907, Taiwan}
\author{G. Y. Guo}
\email{gyguo@phys.ntu.edu.tw}
\affiliation{Department of Physics, National Taiwan University, Taipei 106,  
Taiwan\\
and Department of Physics, Chinese University of Hong Kong, Shatin, N. T., Hong Kong}
\date{\today}
\begin{abstract}
We have theoretically designed three families of the half-metallic (HM) 
antiferromagnets (AFM), namely, La$A$VOsO$_6$, La$A$MoTcO$_6$ and La$A$MoReO$_6$
($A$ = Ca, Sr, Ba), based on a systematic {\it ab initio} study of the
ordered double perovskites La$ABB'$O$_6$ with the possible $B$ and $B'$ pairs
from all the 3$d$, 4$d$ and 5$d$ transtion metal elements being considered. 
Electronic structure calculations based on first-principles density-functional 
theory with generalized gradient approximation (GGA) for more than sixty double
perovskites LaCa$BB'$O$_6$ have been performed using the all-electron 
full-potential linearized augmented-plane-wave method. 
The found HM-AFM state in these materials survives the full {\it ab initio}
lattice constant and atomic position optimizations which were carried out 
using frozen-core full potential projector augmented wave method.
It is found that the HM-AFM properties predicted previously in some of 
the double perovskites would disappear after the full structural optimizations. 
The AFM is attributed to both the superexchange mechanism and the generalized double 
exchange mechanism via the $B$ ($t_{2g}$) - O (2$p_{\pi}$) - $B'$ ($t_{2g}$) 
coupling and the latter is also believed to be the origin of the HM.
Finally, in our search for the HM-AFMs, 
we find La$A$CrTcO$_6$ and La$A$CrReO$_6$ to be AFM insulators of an unconventional
type in the sense that the two antiferromagnetic coupled ions consist of two
different elements and that the two spin-resolved densities of states are no longer
the same. It is hoped that our interesting predictions would stimulate further
experimental searches for the HM-AFMs which have so far been unsuccessful. 

\pacs{71.20.-b, 75.10.Lp, 75.50.Ee, 75.80.+w}
\end{abstract}

\maketitle

\section{INTRODUCTION}
Half-metallic (HM) ferromagnets (FM) was first discovered by 
de Groot, {\it et al.}~\cite{gro83}, based on their band structure calculations 
for magnetic semi-Heusler compounds NiMnSb and PtMnSb. Shortly afterwards,
other magnetic materials such as Fe$_3$O$_4$~\cite{yan84} and CrO$_2$~\cite{sch86} were also
found to be half-metals. Half-metallic materials are 
characterized by the coexistence of metallic behaviour for one electron 
spin and insulating behaviour for the other. Their electronic density 
of states is completely spin polarized at the Fermi level, and the conductivity 
is dominated by these metallic single-spin charge carriers. 
Therefore, half-metallic materials offer potential technological applications 
such as a single-spin electron 
source, and high-efficiency magnetic sensors~\cite{par98,kob98,pic01}.

In this work, we search for another kind of HM materials with first-principles 
calculations. In a HM material, the spin magnetic moment per unit cell is quantized,
i.e., an integer number times Bohr magneton ($\mu_B$).~\cite{gro83,kob98,pic01,jen03} 
It may occur that this integer 
for some HM materials is zero. This situation has been called half-matellic 
antiferromagnetism.~\cite{leu95,pic96,rud98,pic01} 
Most properties such as full spin-polarized 
conduction electrons, zero spin susceptibility, and no Stoner continuum, of these 
HM antiferromagnts (AFM) are the same as  
those of the HM-FMs discussed above~\cite{pic01}. However, there is one important difference: a HM-AFM produces 
no macroscopic magnetic field. Therefore, a HM-AF could support 100\% spin polarized charge 
transport without any net magnetization. 
Furthermore, since there is no symmetry 
operation (translation plus spin flip) that connects its spin-up and spin-down 
bands, a HM-AFM is qualitatively different from a conventional AFM such as bcc Cr. 
Because of their unique properties mentioned above, HM-AFMs have recently attracted 
considerable attention. Furthermore, they could be used as, e.g.,  
a probe of the spin-polarized 
scanning tunneling microscope without perturbing the spin character of samples. 
The HM-AFMs are expected to play a vital role in the future spintronic 
devices which utilize the spin polarization of the carriers.

The first HM-AFM was proposed by van Leuken and de Groot~\cite{leu95} on the 
basis of the Heusler compound V$_7$MnFe$_8$Sb$_7$In. Due to the 
complexity of this material, considerable effort has been spent to come 
up with other candidate materials, which may be easier to synthesize. In particular,
based on extensive first-principles band structure calculations for 
double-perovskite-structutre oxides
La$_2BB'$O$_6$ where $B$ and $B'$ are transition metal ions, Pickett~\cite{pic98} 
proposed the cubic double perovskite La$_2$VMnO$_6$ to be a promising candidate for a 
HM-AFM. In this case, V and Mn have antiferromagnetically aligned magnetic
moments and they exactly cancel each other. This finding
was borned out by the assumption that both Mn and V ions are trivalent and more
importantly, all Mn$^{3+}$ ions are in a low spin state 
of $S=1$ ($t_{2g}^3\uparrow t_{2g}^1\downarrow$).
More recently, Park, {\it et al.}~\cite{par01} 
suggested thiospinel systems Mn(CrV)S$_4$ and
Fe$_{0.5}$Cu$_{0.5}$(V$_{0.5}$Ti$_{1.5}$)S$_4$ as possible HM-AFMs, 
based on their electronic structure studies. In Ref. ~\onlinecite{par02},
the mixed-cation double perovskites La$A'$VRuO$_6$ ($A'$ = Ca, Sr, and Ba)
were suggested to be candidates for HM-AFs. 
Androulakis, {\it et al.}~\cite{and02} have synthesized La$_2$VMnO$_6$ samples which, however, 
has cubic, partially ordered double perovskite structure. Furthermore, it exhibits 
ferrimagnetic (FIM) behavior rather than antiferromagnetic one, with Mn and V being trivalent 
and with Mn being in high spin state.
Therefore, to date, there has been no successful experimental 
realization of the HM-AFM, though it has been speculated in Refs. ~\onlinecite{par02,pic98} that 
ordered mixed-cation double perovskites $AA'BB'$O$_6$ would be promising 
candidates for the HM-AFM. 

Motivated by the theoretical and experimental works mentioned above, we have taken
a rather thorough search for HM-AFMs among the mixed-cation double perovskites La$ABB'$O$_6$.
Unlike the previous theoretical effort~\cite{pic98,par02}, we have not only performed 
electronic structure calculations for a fixed crystal structure but also carried 
out full structural optimizations
including lattice constants and atomic position relaxations.
We believe that the structural optimizations would be important.
In particular, though La$_2$VMnO$_6$ was predicted to be a HM-AFM in Ref. ~\onlinecite{pic98},
it was found to be ferrimagnetic in a partly ordered double perovskite structure~\cite{and02}.
We have explored a variety of pairs $BB'$ with $B$ and $B'$ from 3$d$ (V, Cr, Co, Ni), 4$d$ 
(Mo, Tc, Ru, Rh) and 5$d$ (W, Re, Os, Ir) transition metals. 
Fortunately, we find double perovskites La$A$VOsO$_6$ and La$A$Mo$Y$O$_6$ 
($Y$ = Tc, Re) to be robust HM-AFMs.  

\section{THEORY AND COMPUTATIONAL DETAILS}
Our {\it ab initio} theoretical search was based on first-principles calculations
within density-functional theory (DFT)~\cite{hoh65} with the local density approximation
(LDA) plus the generalized gradient corrections (GGA)~\cite{per96}.
We used the highly accurate all-electron full-potential linearized augmented plane wave 
(FLAPW) method~\cite{and75,koe75} as implemented in the WIEN2k package~\cite{bla02},
to calculate the total energy, electronic band structure and
magnetic properties for the materials with fixed structural parameters. 
The FLAPW method makes no shape approximations to the electron density 
or potential and retains high variational freedom in all regions, and hence it is 
well suited to open crystal structures with low site-symmetries such as 
those considered here. The wave function,
charge density, and potential were expanded in term of the spherical harmonics
inside the muffin-tin spheres. The cut off angular momentum ($L_{max}$) of 10
used for the wave function and of 6 used for the charge density and potential
are sufficient for accurate total-energy calculations~\cite{bla02}. The wave
function outside the muffin-tin spheres was expanded in terms of the augmented
plane waves. A large number of augmented plane waves (about 115 per atom)
(i.e., $R_{mt}K_{max} = 6$) were included in the present calculations.
The set of basis functions was supplemented with 
local orbitals for additional flexibility in representing the valence V 3$d$ states, 
Ru 4$d$ states, Os 5$d$ states and La 4$f$ states, as well as the semicore 
Ca (or Sr, Ba) 3$s$, 3$p$, La 5$s$, 5$p$, O 2$s$ states.
The muffin-tin sphere radii used are 2.5 a.u. for La and Ca (or Sr, Ba), 
and 2.0 a.u. for V, Ru, Os, Mo, Cr, Tc and Re, 1.4 a.u. for O. 
The improved tetrahedron method
is used for the Brillouin-zone integration~\cite{blo94a}.
We used 140 $k$-points in the irreducible 
Brillouin zone wedge, which correspond to 1500 $k$-points in the first Brillouin-zone. 

For the structural optimization calculations to determine theoretical
lattice constants and atomic positions, we used the faster frozen-core full-potential
projector augmented wave method (PAW)~\cite{blo94b} as implemented in
the VASP package~\cite{kre96}. A cutoff energy of 450 eV for plane waves
is used. A 8$\times$8$\times$6 Monkhorst-Pack $k$-point grid in the Brillouin
zone was used, which correspond to 30 $k$-points in the irreducible
Brillouin-zone wedge. Their atomic positions and lattice constants were 
fully relaxed by a conjugate gradient technique. Theoretical equilibrium 
structures were obtained when the forces acting on all the atoms and the 
stresses were less than 0.05 eV/\AA$ $
and 1.5 kBar, respectively.


\section{CRYSTAL STRUCTURE AND THEORETICAL SEARCH STRATEGY}
We consider La$ABB'$O$_6$ ($A$ = Ca, Sr, Ba) in an ordered double perovskite
structure [space group P4/nmm (No. 129)], as shown schematically in Fig. 1.
This structure can be regarded as a combination of
cubic La$B$O$_3$ and $AB'$O$_3$ perovskites in a superlattice 
having $B$ and $B'$ layers stacked 
along the [111] direction, as in La$_2$CrFeO$_6$. Transition metal ions 
$B$ and $B'$ can couple to each other either ferromagnetically or antiferromagnetically
in the [111] directions. Each $B$ ($B'$) ion has six $B'$ ($B$) neighbors.
La$B$O$_3$ and $AB'$O$_3$ perovskites 
can also be stacked along the [001] direction, resulting in a double
perovskite structure with the space group P4mm (No. 99), as has been
considered by Park, {\it et al.}~\cite{par02}. 
In this case, each $B$ ion has four $B$ ion and two $B'$ ion neighbors, while
each $B'$ ion has four $B'$ ion and two $B$ ion neighbors.
We have performed electronic structure calculations for La$A$VRuO$_6$ and La$A$VOsO$_6$
in both P4/nmm and P4mm structures. We find, however, that La$A$VRuO$_6$ and La$A$VOsO$_6$
in the [001] stacked structure are nonmagnetic and ferrimagnetic, respectively.
Furthermore, the calculated total energy for the [001] stacked structure 
is about 0.25$\sim$0.35 eV/f.u. higher than that of the [111] stacked structure.
This shows that the [111] stacked structure is more stable and thus we will not consider
the [001] stacked structure further.

\begin{figure}
\caption{\label{fig:fig1} (Color online, the file size of this figure is too big to be included) The crystal structure of an ordered double perovskite
(P4/nmm).}
\end{figure}

The [111] stacked structure (P4/nmm) contains twenty atoms per unit cell, i.e.,
two chemical formula units (f.u.). It has a tetragonal symmetry with lattice
constants $a$ and $c$. The atomic positions of two La and two $A$ atoms are
($1/4,3/4,0$), ($3/4,1/4,0$) and ($1/4,3/4,1/2$), ($3/4,1/4,1/2$), respectively.
The atomic positions of two $B$ and two $B'$ atoms are
($1/4,1/4,1/4+\delta$), ($3/4,3/4,3/4-\delta$) and ($3/4,3/4,1/4+\delta'$), 
($1/4,1/4,3/4-\delta'$), respectively. The O atoms are divided into three different
types. The atomic positions of two type one and two type two O atoms are O$_1$: 
($1/4,1/4,z_1$), ($3/4,3/4,1 - z_1$) and  O$_2$: ($1/4,1/4,1/2-z_2$),
($3/4,3/4,1/2+z_2$), respectively, while the atomic positions of eight type three
O atoms are O$_3$: ($1/2-x,1/2-x,1/4-z_3$), ($x,x,1/4-z_3$), ($x,1/2-x,1/4-z_3$),
($1/2-x,x,1/4-z_3$), ($1/2+x,1-x,3/4+z_3$), ($1-x,1/2+x,3/4+z_3$), ($1-x,1-x,3/4+z_3$),
($1/2+x,1/2+x,3/4+z_3$). Therefore, in this structure, apart from lattice constants
$a$ and $c$, there are six more parameters to optimize, namely, $\delta$, $\delta'$,
$x$, $z_1$, $z_2$, $z_3$. 

\begin{figure}
\includegraphics[width=8cm]{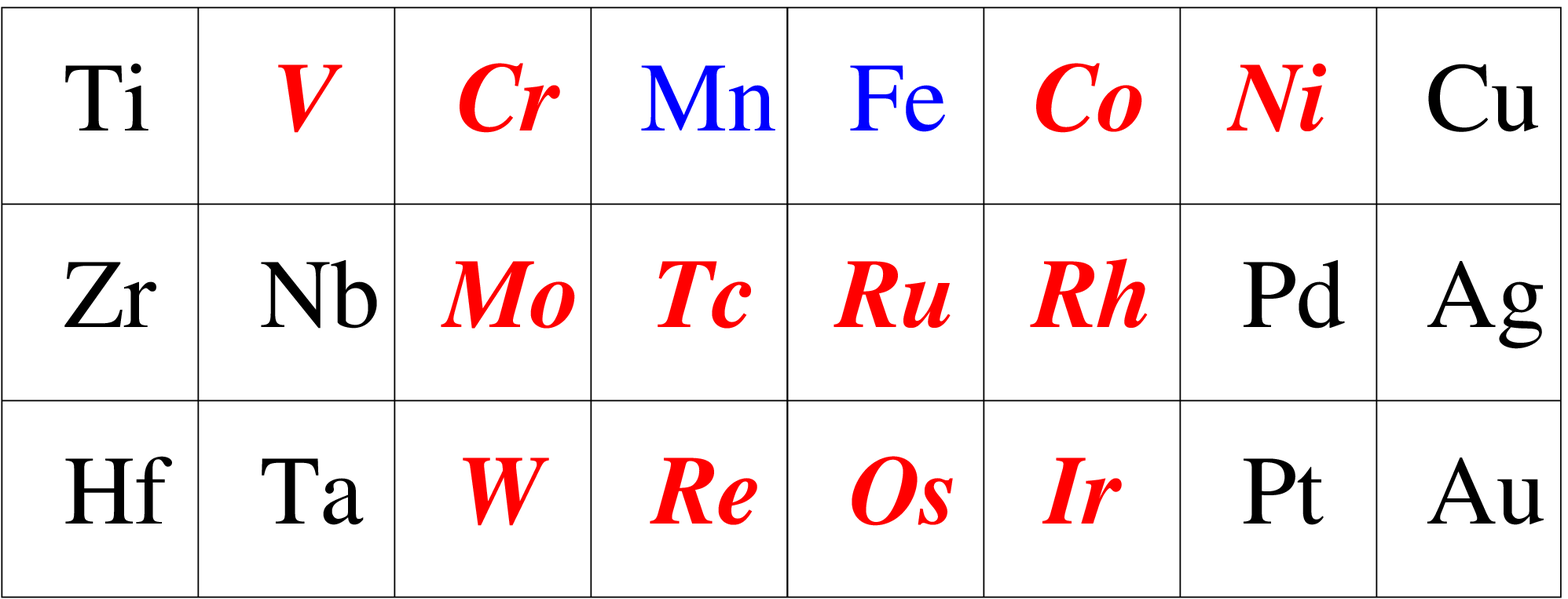}
\caption{\label{fig:fig2} (Color online) Part of the Periodic Table showing 3$d$, 4$d$ and 5$d$ 
transition metals. Bold italic characters denote the twelve elements that have been
selected as the $BB'$ pairs for La$ABB'$O$_6$ in this work.}
\end{figure}

For the chemical formula La$ABB'$O$_6$, $B$ and $B'$ can be an any pair from the twenty-four transition
metal elements, as shown in Fig. 2, and $A$ is one of the three alkali earth elements Ca, Sr, and Ba. 
Therefore, although we have settled down on the ordered double perovskite structure, there are still 
3$\times C_2^{24} = 828$ possible compounds to consider. This is an impossible task. Nevertheless, 
our preliminary calculations and also the existing literature tell us that Ti, Cu, Zr, Nb, Pd, Ag, Hf,
Ta, Pt and Au in the double perovskite structure are usually not magnetic and thus, they can be excluded from our
further consideration. Furthermore, we note that in the double perovskite structure, Mn and Fe are almost always 
in a high spin state (see, e.g., Refs. \onlinecite{pic96,par01,jen03}). 
Consequantly, it is not possible to find a high spin state element
from the remaining twelve elements to match Mn and Fe, and hence Mn and Fe are excluded from 
further consideration too. Therefore, in our first stage of search for
HM-AFs, we consider all the possible $BB'$ pairs from only the twelve transition metal elements as denoted
by bold italic characters in Fig. 2. Furthermore, we initially confine ourselves to $A =$ Ca. Therefore,
in our first stage of search, we have performed self-consistent electronic structure
calculations only for sixty-six ($c_2^{12} = 66)$) LaCa$BB'$O$_6$ compounds in the ordered double perovskite structure with
fixed lattice constants $a = 5.492$ \AA, $ c = \sqrt{2}a$ taken from Ref. \onlinecite{par02}. When we find any
LaCa$BB'$O$_6$ to be a HM-AFM, we replace Ca with Sr and Ba and perform further self-consistent electronic
structure calculations.

Fortunately, we find from our electronic structure calculations for sixty-six 
LaCa$BB'$O$_6$ compounds, that six $B$ and $B'$ pairs, namely,
VRu, VOs, MoTc, MoRe, NiTc, NiRe, would give rise to HM-AF
LaCa$BB'$O$_6$ compounds.  We note that previously, Park, {\it et al.}~\cite{par02} 
theoretically predict La$A$VRuO$_6$ to be HM-AFMs.
However, we performed both volume and
atomic position optimization calculations for these systems and found,
unfortunately, that 
the HM-AFM properties predicted in Ref. \onlinecite{par02}, disappear.
This perhaps explains why the attempt to synthesize the HM-AFM La$A$VRuO$_6$
by Liu, {\it et al.},\cite{liu02} was unsuccessful.
Therefore, in our second stage of search, we carried out full lattice constant and
atomic position relaxation calculations for all the possible HM-AFM La$ABB'$O$_6$ compounds
found in the first stage of search. We find fortunately that among the possible HM-AFM La$ABB'$O$_6$ compounds,
La$A$VOsO$_6$, La$A$MoTcO$_6$ and La$A$MoReO$_6$ stay as the stable HM-AFMs.

\begin{figure}
\includegraphics[width=8cm]{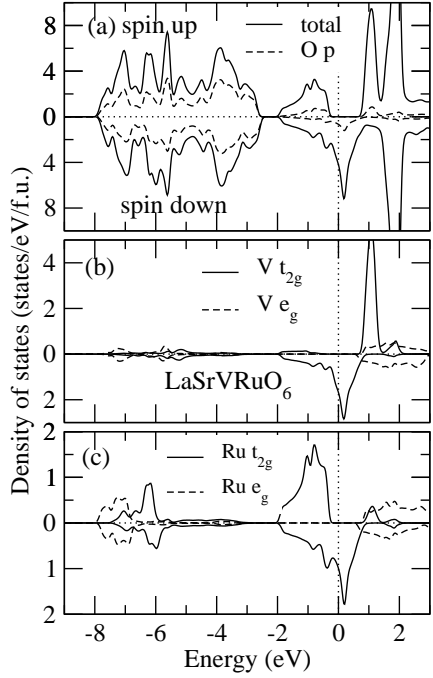}
\caption{\label{fig:fig3} Total and orbital-decomposed density of states of
LaSrVRuO$_6$ in the ideal P4/nmm structure (no atomic position relaxation, 
see text and Table I).} \end{figure}

\begin{table}
\caption{Calculated physical properties of La$A$VRuO$_6$ in the ideal P4/nmm structure
($\delta = \delta'= x = z_1 = z_2 = z_3 = 0$).}
\begin{ruledtabular}
\begin{tabular}{cccccccc}
\multicolumn{2}{c}{$A = $}  & \multicolumn{2}{c}{Ca} & \multicolumn{2}{c}{Sr} & \multicolumn{2}{c}{Ba}\\ \hline
\multicolumn{2}{c}{magnetic states} & FM/NM  &  AF & FM/NM &  AF & FM/NM &  AF\\ \hline
\multicolumn{2}{c}{$a$ (\AA)}  &5.510 & 5.517 &5.545 & 5.553& 5.604 & 5.612\\
\multicolumn{2}{c}{$c/a$}  &\multicolumn{2}{c}{$\sqrt{2}$} & \multicolumn{2}{c}{$\sqrt{2}$} & \multicolumn{2}{c}{$\sqrt{2}$} \\
\multicolumn{2}{c}{$V$ (\AA$^3$/f.u.)} & 118.31& 118.72& 120.57 & 121.05 & 124.44 & 125.00 \\ \hline
\multicolumn{2}{c}{$m_V$ ($\mu_B$)} &  0.039 &  -0.949 &  0.017 &  -0.967    &  0.006 &  -0.992 \\
  \multicolumn{2}{c}{$m_{Ru}$ ($\mu_B$)} & 0.074 & 0.730 &  0.028 &  0.738    &  0.016 & 0.744 \\
  \multicolumn{2}{c}{$m_t$ ($\mu_B$/f.u.)}  &  0.165 &   0.0  &  0.065 & 0.0  &   0.033 &  0.0 \\ \hline
  $N(E_F)$  & $\uparrow$ & 2.563 & 0.0  &  2.575 & 0.0  &  2.358  & 0.0 \\
  (states/eV/f.u.)& $\downarrow$ & 2.491 & 4.064    &  2.640 &  4.205  &  2.460 &   4.414 \\ \hline
  gap (eV) & $\uparrow$ &  & 0.74 &  & 0.79  &  & 0.79 \\
  \multicolumn{2}{c}{$\Delta E^{AF-FM}$ (eV/f.u.)} &\multicolumn{2}{c}{-0.102}  &   \multicolumn{2}{c}{-0.103}  & \multicolumn{2}{c}{-0.090} \\ 
\end{tabular}
\end{ruledtabular}
\end{table}

\section{RESULTS and DISCUSSION}
                                                                                                 
\begin{table}
\caption{Calculated physical properties of La$A$VOsO$_6$ in the ideal P4/nmm structure
($\delta = \delta'= x = z_1 = z_2 = z_3 = 0$).}
\begin{ruledtabular}
\begin{tabular}{cccccccc}
\multicolumn{2}{c}{$A = $}  & \multicolumn{2}{c}{Ca} & \multicolumn{2}{c}{Sr} & \multicolumn{2}{c}{Ba}\\ \hline
\multicolumn{2}{c}{magnetic states} & FM/NM  &  AF & FM/NM &  AF & FM/NM &  AF\\ \hline
\multicolumn{2}{c}{$a$ (\AA)}  & 5.522 & 5.546 & 5.557 & 5.577& 5.607 & 5.635\\
\multicolumn{2}{c}{$c/a$}  &\multicolumn{2}{c}{$\sqrt{2}$} & \multicolumn{2}{c}{$\sqrt{2}$} & \multicolumn{2}{c}{$\sqrt{2}$} \\
\multicolumn{2}{c}{$V$ (\AA$^3$/f.u.)} & 119.05 & 120.61& 121.36 & 122.67& 124.62 & 126.53 \\ \hline
\multicolumn{2}{c}{$m_V$ ($\mu_B$)} &  0.0 &  -1.254 &  0.0 &   -1.274    &  0.0 &  -1.305 \\
  \multicolumn{2}{c}{$m_{Os}$ ($\mu_B$)} & 0.0 & 0.890 &  0.0 &  0.899    &  0.0 & 0.910 \\
  \multicolumn{2}{c}{$m_t$ ($\mu_B$/f.u.)}  &  0.0   &   0.0  &  0.0 & 0.0  &   0.0 &  0.0 \\
\hline
  $N(E_F)$  & $\uparrow$ & 3.420 & 0.0  &  3.462& 0.0 &  2.527  & 0.0 \\
  (states/eV/f.u.)& $\downarrow$ & 3.420 & 3.798    &  3.462&  3.883  &  2.527 &   4.023 \\ \hline
  gap (eV) & $\uparrow$ &  & 0.82 &  & 0.87  &  & 0.98 \\
  \multicolumn{2}{c}{$\Delta E^{AF-FM} (eV/f.u.)$} &\multicolumn{2}{c}{-0.231}  &   \multicolumn{2}{c}{-0.236}  & \multicolumn{2}{c}{-0.237} \\
\end{tabular}
\end{ruledtabular}
\end{table}

\subsection{Half-metallic antiferromagnets from the initial search}

In the first round of search for the HM-AFMs, we have performed $ab$ $initio$ electronic structure
calculations for sixty-six LaCa$BB'$O$_6$ compounds in the ideal P4/nmm structure
(i.e., $\delta = \delta'= x = z_1 = z_2 = z_3 = 0$) with the lattice constants
 $a = 5.492$ \AA, $ c = \sqrt{2}a$ taken from Ref. ~\onlinecite{par02}, as mentioned before. 
Remarkably, we found LaCaV$Y$O$_6$, LaCaMo$Y'$O$_6$ and LaCaNi$Y'$O$_6$ ($Y =$ Ru, OS; 
$Y'$ = Tc, Re) to be HM-AFMs. We then replaced Ca in these compounds with either
Sr or Ba and performed further electronic structure calculations. We found again that
the resultant compounds are HM-AFMs. This result could be expected because
$A$-site atoms in $AB$O$_3$
perovskite are known to behave like a carrier reservoir and a volume conserver. The interaction
between an $A$ atom and its neighboring atoms is usually very weak such that 
many $AB$O$_3$s with different $A$s, have similar electronic properties.

In the ionic picture, the atoms in the ordered double perovskites have the nominal
valence states as La$^{3+}A^{2+}$($BB'$)$^{7+}$O$_6^{2-}$. Therefore,
in, e.g., La$A$VRuO$_6$, the transition metal atoms $B$ and $B'$ can have
valence configurations of V$^{3+}$(3$d^2$) and Ru$^{4+}$(4$d^4$), and 
the antiferromagnetic coupling of the high spin V$^{3+}$ ($t_{2g}^2\uparrow$, S=1) 
and the low spin Ru$^{4+}$ ($t_{2g}^3\uparrow t_{2g}^1\downarrow$, S=1) 
states would give rise to a zero total magnetic moment, i.e., a AFM state,
Alternatively, the transition metal atoms $B$ and $B'$ can have
valence configurations of V$^{4+}$(3$d^1$) and Ru$^{3+}$(4$d^5$), and 
the antiferromagnetic coupling of the spin V$^{4+}$ ($t_{2g}^1\uparrow$, S=1/2)
and the low spin Ru$^{3+}$ ($t_{2g}^3\uparrow t_{2g}^2\downarrow$, S=1/2) states. 
Of course, in the real materials, the above simple ionic model would be modified
because of hybridization between V $3d$ and O $2p$ orbitals and also between
Ru $4d$ and O $2p$ orbitals. The calculated total and orbital-decomposed density 
of states (DOS) of LaSrVRuO$_6$ in the AFM state are shown in Fig. 3. 
Clearly, LaSrVRuO$_6$ is half-metallic with a band gap of about 0.8 eV on
the spin-up channel. The electronic structure consists of the O $2p$ dominant
lower valence band between 2.5 and 8.0 eV below the Fermi level ($E_F$), the spin-up
Ru $t_{2g}$ dominant upper valence band between 0.3 and 2.0 eV below $E_F$
and spin-down Ru $t_{2g}$-V $t_{2g}$ hybridized conduction band between 2.0 eV 
below $E_F$ and $\sim$1.0 eV above $E_F$ as well as the spin-up V $t_{2g}$ dominant,
Ru $e_g$ dominant and V $e_g$ dominant upper conduction bands about 0.8 eV
above $E_F$ (see Fig. 3). The calculated DOS spectra are very similar to
the ones reported in Ref. ~\onlinecite{par02}. 
Our calculated occupation numbers are 0.72 e (spin-up) and 1.68 e (spin-down)
at the V-site, and 2.72 e (spin-up) and 1.98 e (spin-down) on the Ru-site.
Therefore, the calculated charge configurations are V$^{2.6+}$ (3$d^{2.4}$) and 
Ru$^{3.3+}$ (4$d^{4.7}$), suggesting that the valence configurations of 
V$^{3+}$/Ru$^{4+}$ and V$^{4+}$/Ru$^{3+}$ are perhaps nearly
degenerate and hence highly mixed, as pointed out before~\cite{par02}.
This notion is further supported by the fact that the calculated local spin
magnetic moments of V and Ru are not integer numbers of 2 or 1 but less than 1.0
(see Table I).

Similarly, we can understand the formation of the AFM state in the other
HM-AFM compouds in the ionic picture. As for La$A$VRuO$_6$, in La$A$VOsO$_6$, 
the transition metal atoms $B$ and $B'$ can have valence configurations 
of V$^{3+}$(3$d^2$) and Os$^{4+}$(5$d^4$), and
the antiferromagnetic coupling of the high spin V$^{3+}$ ($t_{2g}^2\uparrow$, S=1)
and the low spin Os$^{4+}$ ($t_{2g}^3\uparrow t_{2g}^1\downarrow$, S=1)
states would give rise to a zero total magnetic moment.
In La$A$MoTc(Re)O$_6$, the transition metal atoms $B$ and $B'$ may have 
valence configurations of Mo$^{3+}$(4$d^3$) and Tc$^{4+}$(4$d^3$) [Re$^{4+}$(5$d^3$)]
with the antiferromagnetic coupling of the low spin Mo$^{3+}$ 
($t_{2g}^2\uparrow t_{2g}^1\downarrow$, S=1/2)
and the low spin Tc(Re)$^{4+}$ ($t_{2g}^2\uparrow t_{2g}^1\downarrow$, S=1/2)
states. In La$A$NiTc(Re)O$_6$, the transition metal atoms $B$ and $B'$ may have 
valence configurations of Ni$^{2+}$(3$d^8$) and Tc$^{5+}$(4$d^2$) [Re$^{5+}$(5$d^2$)]
with the antiferromagnetic coupling of the high spin Ni$^{2+}$ ($t_{2g}^3\uparrow t_{2g}^3\downarrow e_g^2\uparrow$, S=1)
and the high spin Tc(Re)$^{5+}$ ($t_{2g}^2\uparrow t_{2g}^0\downarrow$, S=1)
states.

\begin{table}
\caption{Calculated physical properties of La$A$MoTcO$_6$ in the ideal P4/nmm structure
($\delta = \delta'= x = z_1 = z_2 = z_3 = 0$).}
\begin{ruledtabular}
\begin{tabular}{cccccccc}
\multicolumn{2}{c}{$A = $}  & \multicolumn{2}{c}{Ca} & \multicolumn{2}{c}{Sr} & \multicolumn{2}{c}{Ba}\\ \hline
\multicolumn{2}{c}{magnetic states} & FM/NM  &  AF & FM/NM &  AF & FM/NM &  AF\\ \hline
\multicolumn{2}{c}{$a$ (\AA)}  & 5.629 & 5.640&5.657 & 5.667& 5.699 & 5.717\\
\multicolumn{2}{c}{$c/a$}  &\multicolumn{2}{c}{$\sqrt{2}$} & \multicolumn{2}{c}{$\sqrt{2}$} & \multicolumn{2}{c}{$\sqrt{2}$} \\
\multicolumn{2}{c}{$V$ (\AA$^3$/f.u.)} &126.13 & 126.87&127.99 & 128.66&130.91 & 132.16 \\ \hline
\multicolumn{2}{c}{$m_{Mo}$ ($\mu_B$)} &  0.0 &  -1.007 &  0.0 &  -1.030    &  0.0 & -1.064 \\
  \multicolumn{2}{c}{$m_{Tc}$ ($\mu_B$)} & 0.0 & 1.112 &  0.0 &   1.133    &  0.0 &  1.164 \\
  \multicolumn{2}{c}{$m_t$ ($\mu_B$/f.u.)}  &  0.0 &   0.0  &  0.0 & 0.0  &   0.0 &  0.0 \\
\hline
  $N(E_F)$  & $\uparrow$ & 3.168 & 0.0  &  3.225 & 0.0  &  3.321  & 0.0 \\
  (states/eV/f.u.)& $\downarrow$ & 3.168 & 3.160    &  3.225 &  3.184  &  3.321 &   3.271 \\ \hline
  gap (eV) & $\uparrow$ & &0.44 & & 0.49  &  &0.60 \\
  \multicolumn{2}{c}{$\Delta E^{AF-FM} (eV/f.u.)$} &\multicolumn{2}{c}{-0.146}  &   \multicolumn{2}{c}{-0.159}  & \multicolumn{2}{c}{-0.171} \\
\end{tabular}
\end{ruledtabular}
\end{table}

To see whether the HM-AFM is the stable state in these six compounds, we performed the FM
electronic structure calculations. Unfortunately, we find that the total energy of the HM-AFM state
in LaCaNi$Y'$O$_6$ ($Y'$ = Tc, Re) is substantially higher than that of the FM state, by
0.19 and 0.18 eV/formula unit (f.u.), respectively. This suggests that the HM-AFM state in these
two systems is not stable and hence we will not consider them further.
We then performed volume relaxation calculations for La$A$V$Y$O$_6$ and La$A$Mo$Y'$O$_6$
($A = $ Ca, Sr, Ba; $Y =$ Ru, Os; $Y'$ = Tc, Re) starting with both the FM and AFM states.
We find all these systems to remain as a stable HM-AFM after the volume relaxation.
The calculated electronic and magnetic properties of these four families of the double perovskites
are listed in Tables I-IV, respectively.

\begin{table}
\caption{Calculated physical properties of La$A$MoReO$_6$ in the ideal P4/nmm structure
($\delta = \delta'= x = z_1 = z_2 = z_3 = 0$).}
\begin{ruledtabular}
\begin{tabular}{cccccccc}
\multicolumn{2}{c}{$A = $}  & \multicolumn{2}{c}{Ca} & \multicolumn{2}{c}{Sr} & \multicolumn{2}{c}{Ba}\\ \hline
\multicolumn{2}{c}{magnetic states} & FM/NM  &  AF & FM/NM &  AF & FM/NM &  AF\\ \hline
\multicolumn{2}{c}{$a$ (\AA)}  &5.640 & 5.656&5.669 & 5.687& 5.739 & 5.739\\
\multicolumn{2}{c}{$c/a$}  &\multicolumn{2}{c}{$\sqrt{2}$} & \multicolumn{2}{c}{$\sqrt{2}$} & \multicolumn{2}{c}{$\sqrt{2}$} \\
\multicolumn{2}{c}{$V$ (\AA$^3$/f.u.)} &126.89 & 127.93 &128.85 & 130.06&133.67 & 133.64 \\ \hline
\multicolumn{2}{c}{$m_{Mo}$ ($\mu_B$)} &  0.0 &  -1.138 &  0.0 &  -1.175    &  0.0 & -1.210 \\
  \multicolumn{2}{c}{$m_{Re}$ ($\mu_B$)} & 0.0 &  1.100 &  0.0 &   1.133    &  0.0 &  1.161 \\
  \multicolumn{2}{c}{$m_t$ ($\mu_B$/f.u.)}  &  0.0 &   0.0  &  0.0 & 0.0  &   0.0 &  0.0 \\
\hline
  $N(E_F)$  & $\uparrow$ & 2.925 & 0.0  &  3.003 & 0.0  &  3.175  & 0.0\\
  (states/eV/f.u.)& $\downarrow$ & 2.925 & 2.190  &  3.003 &  1.853  &  3.175 &   1.423 \\ \hline
  gap (eV) & $\uparrow$ & & 0.16 &  &0.27  &  &0.35 \\
  \multicolumn{2}{c}{$\Delta E^{AF-FM} (eV/f.u.)$} &\multicolumn{2}{c}{-0.148}  &   \multicolumn{2}{c}{-0.172}  & \multicolumn{2}{c}{-0.214} \\
\end{tabular}
\end{ruledtabular}
\end{table}

Tables I-IV show clearly that the total energy of the HM-AFM state in these four families of the
double perovskites is significantly lower than that of the corresponding FM or nonmagnetic (NM)
state. The energy difference is larger than 0.1 eV/f.u. and can be as large as $\sim$ 0.24 eV/f.u.,
e.g., in LaBaVOsO$_6$ (Table II). The insulating gap in the spin-up band structure is rather large 
and can be up to 1.0 eV as in LaBaVOsO$_6$ (Table II). It is also interesting to note that no 
strong FM solution can be stabilized in all the four families of the compounds, and hence these
compounds have either a weak FM or NM metastable state (Tables I-IV). Nonetheless, the local magnetic
moments in the HM-AFM state are not large either, all being in the order of 1.0 $\mu_B$/atom. 

\subsection{Effects of structural optimization}

In our second stage of search, we carried out full 
structural optimization (i.e., both lattice constant and atomic position relaxations) 
calculations for all the possible HM-AFM
La$ABB'$O$_6$ compounds
described in the previous subsession. Interestingly,  we find that 
La$A$VOsO$_6$, La$A$MoTcO$_6$ and La$A$MoReO$_6$ ($A =$ Ca, Sr, and Ba) 
remain to be the stable HM-AFMs,
although La$A$VRuO$_6$ become nonmagnetic after the full structural optimization.
The equilibrium lattice constants and atomic positions as well as the stability
of the HM-AFM of these three families of the double perovskites are listed
in Table V. Clearly, the HM-AFM state is stable over the FM or NM state in these
compounds, and the size of the energy difference between the HM-AFM and
FM/NM states is similar to the one before the full structural optimization
(see Tables I-IV).
The full structural optimization further lower the total energy of these compounds
rather significantly. The most pronounced decrease in the total energy occurs
in the Ba-based double perovskites. For example, the decrease is about 0.2$\sim$0.3
eV/f.u. for LaBaMo$Y$O$_6$ ($Y =$ Tc, Re) and about 0.3$\sim$0.4 eV/f.u. for
LaBaVOsO$_6$. This may be attribued to the fact that the ionic
radius of Ba is significantly larger than that of Ca. This notion is further supported by
the fact that the equilibrium unit cell volume increases as one replaces Ca with Sr and
then with Ba (see Table V). The decrease in the total energy
become smaller as one moves from Ba through Sr to Ca. The decrease in the total energy
is about 0.1$\sim$0.15 eV/f.u. for
LaSrMo$Y$O$_6$ and about 0.05$\sim$0.1 eV/f.u. for
LaCaMo$Y$O$_6$ ($Y =$ Tc, Re).

The shape (i.e., the $c/a$ ratio) of the unit cell for the ordered double
perovskites in the HM-AFM state 
does not change much after the full structural optimization (Table V).
In particular, the deviation of the $c/a$ for LaCa$BB'$O$_6$ from the
ideal value of $\sqrt{2}$ is less than 1 \%. This is because La and Ca have
rather similar atomic radii of 3.92 and 4.12 a.u.\cite{skr84}, respectively.
The deviation increases very slightly as Ca is replaced by Sr and
Ba, and the largest deviation of about 1.3 \% occurs in LaBaVOsO$_6$.  
Sr and Ba have larger atomic radii of 4.49 and 4.65, respectively.
However, the internal atomic displacements after the structural optimization
are rather pronounced. For example, the contraction (expansion) of 
the in-plane $B$-O ($B'$-O) bondlength of the $B$O$_6$ ($B'$O$_6$) 
octohedra which is 
described by the parameter $x$, can be as large as 2.7 \% 
(or 0.038 \AA) in LaCaMoTcO$_6$. The in-plane $B$-O-$B'$ bond bending 
which is described by the parameters $z_3$, $\delta$ and $\delta '$, 
is also rather significant. In particular, the $B$-O-$B'$ bond bending 
can be as large as 8$^{\circ}$ (i.e., the angle of the $B$-O-$B'$ bond 
is about 172$^{\circ}$) in LaBaVOsO$_6$. A similar amount of the
bond bending occurs in LaBaMoTcO$_6$ and LaBaMoReO$_6$. This may be
attributed to the fact that the atomic radius of Ba is considerably
larger than that of La. This bond bending becomes smaller when
Ba is replaced by Sr and Ca. Another significant atomic displacement
is the contraction (expansion) of the apical $B$-O ($B'$-O) bond of the
$B$O$_6$ ($B'$O$_6$) octohedra which is described by the parameters
$z_1$, $z_2$, $\delta$ and $\delta '$. For example, the contraction 
of the V-O$_2$ bondlength is about 4.3 \%. Interestingly, this displacement
of the apical O atoms is asymmetrical. Indeed, the V-O$_1$ bond is elongated
slightly, i.e., expands by 0.21 \%. 

Obviously, the full structural optimization has the largest effects on
La$A$VRuO$_6$ in which both the HM-AF and FM states disappear and the
systems become nonmagnetic after the structural optimization, as mentioned
before. The lowering of the total energy due to the structural optimization
is the largest among the four families of the HM-AFM candidates found in
our first stage of search, varying from 0.4 to 0.8 eV/f.u. Likewise,
the contraction of the VO$_6$ octahedra and the expansion of the RuO$_6$
octahedra during the structural optimization are also the largest (see
Tables V and VI). For example, the contraction (expansion) of the V-O (Ru-O) 
bond on
the $xy$ plane is about 3.0 \% and that of the apical V-O$_2$ bond is as large as 6.4 \%.
Furthermore, the $B$-O-$B'$ bond bending is around 7$^{\circ}$ in all the three La$A$VRuO$_6$ compounds.
It is therefore conceivable that the weak AFM and FM state of La$A$VRuO$_6$
found in the ideal P4/nmm structure could not sustain such large structural 
distortions and both of them become nonmagnetic.

\begin{figure}
\includegraphics[width=8cm]{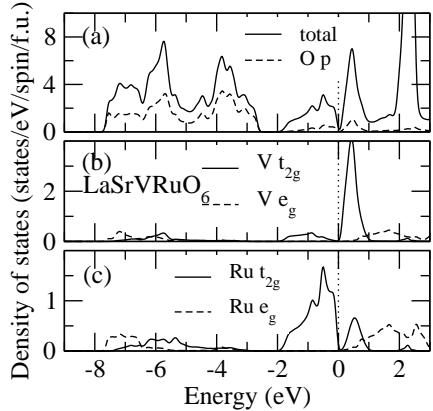}
\caption{\label{fig:fig4} Total and orbital-decomposed density of states of LaSrVRuO$_6$ in the theoretical determined P4/nmm structure (full structural optimization, see text).} \end{figure}

Why the magnetic states disappear in La$A$VRuO$_6$ after the full structural
optimization can perhaps be best understood by examining their electronic
structures. Fig. 4 shows the total and orbital-decomposed densities of states
of LaSrVRuO$_6$ in the fully optimized structure. They are similar to the ones
for LaSrVRuO$_6$ in the ideal P4/nmm structure displayed in Fig. 3, except that
there are not spin-polarized anymore. The electronic structure consists of the O $2p$ 
dominant lower valence band between 2.7 and 7.7 eV below $E_F$, the 
Ru $t_{2g}$ dominant upper valence band between 0.0 and 1.9 eV below $E_F$
and V $t_{2g}$ dominant lower conduction band between 0.0 eV
 and $\sim$1.0 eV above $E_F$ as well as the V $t_{2g}$/Ru $e_g$ dominant
upper conduction bands about 0.5 eV above $E_F$ (see Fig. 4). 
The most striking feature is the pseudo-gap at $E_F$ and LaSrVRuO$_6$ is 
a semimetal with a low DOS of 0.3 states/eV/spin/f.u. at $E_F$.
It is clear that because of the pseudo-gap at $E_F$, La$A$VRuO$_6$ are
now stable against magnetic instabilities and therefore the systems stay
in the nonmagnetic state.~\cite{note1,cyr79} In the nonmagnetic band structures of 
La$A$VOsO$_6$, La$A$MoTcO$_6$ and La$A$MoReO$_6$, there is no such 
pseudo-gap at $E_F$ and hence they retain the HM-AFM state after the
full structural optimization.

\begin{table}
\caption{Theoretical structural parameters and stability of La$ABB'$O$_6$ in the HM-AFM as well
as FM and NM states. The lattice constant $a$ is in the unit of \AA.}
\begin{ruledtabular}
\center{(a) La$A$VOsO$_6$}
\begin{tabular}{ccccccc}
$A = $  & \multicolumn{2}{c}{Ca} & \multicolumn{2}{c}{Sr} & \multicolumn{2}{c}{Ba}\\ \hline
magnetic states & FM/NM  &  AF & FM/NM &  AF & FM/NM &  AF\\ \hline
$a$ ($c/a$)  &5.538 (1.401)& 5.541 (1.405)&5.585(1.395)& 5.580(1.402) & 5.652(1.385) & 5.647(1.396) \\
$\delta (\delta')$ ($10^{-2}$)  & 0.76 (0.39) & 0.71 (0.22) & 0.46 (0.33) &0.39 (0.14) & -0.09 (-0.16) & -0.03 (-0.13)  \\
$ x (z_1)$ ($10^{-2}$)  & 0.60 (0.42) & 0.44 (0.46) & 0.60 (0.35) & 0.42 (0.34) & 0.56 (0.22) & 0.38 (0.22)  \\
$z_2 (z_3)$ ($10^{-2}$)  & 0.79 (0.59) & 0.52 (0.62) & 0.98 (1.14) & 0.70 (1.18) & 1.34 (2.07) &0.90 (1.88)  \\
$V$(\AA$^3$) &119.06& 119.50 & 121.53 & 121.83 & 125.04 & 125.67 \\ \hline

$\Delta E^{AF-FM}$ (eV/f.u.) & \multicolumn{2}{c}{-0.104}  &   \multicolumn{2}{c}{-0.102}  & \multicolumn{2}{c}{-0.099} \\
\end{tabular}
\center{(b) La$A$MoTcO$_6$}
\begin{tabular}{ccccccc}
$A = $  & \multicolumn{2}{c}{Ca} & \multicolumn{2}{c}{Sr} & \multicolumn{2}{c}{Ba}\\ \hline
magnetic states & FM/NM  &  AF & FM/NM &  AF & FM/NM &  AF\\ \hline
$a$ ($c/a$)  &5.643 (1.408)& 5.656(1.408) &5.679(1.404)& 5.689(1.406) & 5.740(1.392) & 5.747 (1.399)\\
$\delta (\delta') $ ($10^{-2}$)  & 0.49 (0.08) & 0.47 (0.07) & 0.14 (-0.01) & 0.15 (-0.11) & -0.34 (0.06) & -0.19 (-0.25)\\
$ x (z_1)$  ($10^{-2}$) & 0.17 (0.26) &0.67 (0.16) & 0.20 (0.15) &0.02 (0.07) & 0.24 (-0.10) & 0.02 (-0.02) \\
$z_2 (z_3)$  ($10^{-2}$) & 0.07 (0.58) &-0.10 (0.03) & 0.26 (1.17) &0.01 (1.36) & 0.62 (1.93) &0.16 (1.98) \\
$V$(\AA$^3$) &126.50& 127.35 & 128.54 & 129.41 & 131.64 & 132.77 \\ \hline
  $\Delta E^{AF-FM}$ (eV/f.u.) & \multicolumn{2}{c}{-0.135}  &   \multicolumn{2}{c}{-0.149}  & \multicolumn{2}{c}{-0.163} \\
\end{tabular}
\center{(c) La$A$MoReO$_6$}
\begin{tabular}{ccccccc}
$A = $  & \multicolumn{2}{c}{Ca} & \multicolumn{2}{c}{Sr} & \multicolumn{2}{c}{Ba}\\ \hline
magnetic states & FM/NM  &  AF & FM/NM &  AF & FM/NM &  AF\\ \hline
$a$($c/a$)  &5.651(1.409) & 5.671(1.408) &5.687(1.408)& 5.703(1.409) & 5.749(1.393) & 5.756 (1.404)\\
$\delta (\delta')$ ($10^{-2}$)  & 0.40 (0.17) & 0.38 (0.14) & 0.20 (0.03) &0.09 (-0.07)&-0.14 (-0.18) & -0.14 (-0.28)\\
$ x (z_1)$  ($10^{-2}$) &-0.08 (-0.03) &-0.18 (-0.07) & -0.08 (-0.05) &-0.18 (-0.11)  &-0.06 (-0.09) & -0.21 (-0.13) \\
$z_2 (z_3)$  ($10^{-2}$) &-0.13 (0.30) & -0.28 (0.57) &-0.10 (0.86) & -0.25 (1.09) &-0.04 (1.80) & -0.27 (1.94) \\
$V$ (\AA$^3$)&127.15& 128.45 & 129.45 & 130.68 & 132.33 & 133.81 \\ \hline
  $\Delta E^{AF-FM}$ (eV/f.u.) & \multicolumn{2}{c}{-0.182}  &   \multicolumn{2}{c}{-0.229}  & \multicolumn{2}{c}{-0.271} \\
\end{tabular}
\end{ruledtabular}

\end{table}

\begin{table}
\caption{Theoretical structural parameters of La$A$VRuO$_6$.}
\begin{ruledtabular}
\begin{tabular}{cccc}
$A = $  & Ca & Sr & Ba\\ \hline
magnetic state & NM  &  NM & NM \\ \hline
$a$ (\AA) &5.517 &5.559& 5.629 \\
$c/a$  &1.409 & 1.403 & 1.393 \\
$\delta (\delta')$ ($10^{-2}$)  & 0.90 (0.40) & 0.36 (0.36) & -0.29 (0.23)  \\
$ x (z_1)$ ($10^{-2}$)  & 0.78 (0.52) & 0.78 (0.38) & 0.78 (0.24)  \\
$z_2 (z_3)$ ($10^{-2}$)  & 0.91 (0.57) & 1.25 (1.19) & 1.70 (2.06)  \\
$V$ (\AA$^3$/f.u.) &118.26& 120.46 &  124.26 
\end{tabular}
\end{ruledtabular}
\end{table}

\subsection{Electronic structure and magnetic properties}
The calculated electronic and magnetic properties of robust
HM-AFM La$ABB'$O$_6$ are listed in Table VII. Clearly, all three families
of La$A$VOsO$_6$, La$A$MoTcO$_6$ and La$A$MoReO$_6$ ($A$ = Ca, Sr, Ba) are
HM-AFMs with an insulating gap in the spin-up channel. The spin-down density of
states at $E_F$ is rather large in La$A$VOsO$_6$ and La$A$MoTcO$_6$, although
it is rather small in La$A$MoReO$_6$. The insulating gap in the spin-up channel
is also rather large (0.5$\sim$1.0 eV) in La$A$VOsO$_6$ and La$A$MoTcO$_6$ but
relatively small (within 0.5 eV) in La$A$MoReO$_6$. 
The calculated local spin magnetic moments in all the double perovskites
are not large and in the order of 1.0 $\mu_B$/atom. This seems to be
consistent with the simple ionic picture described in Subsec. A 
in the case of La$A$MoTc(Re)O$_6$ but at varience with it in the case 
of La$A$VOsO$_6$. The deviation from the simple ionic model is more apparent
in terms of the calculated occupation numbers of the transition metal
$d$ orbitals. For example, in LaSrVOsO$_6$,   
the calculated occupation numbers are 0.72 e (spin-up) and 1.72 e (spin-down)
at the V-site, and 2.40 e (spin-up) and 1.69 e (spin-down) on the Os-site, giving
rise to a charge configuration of V$^{2.6+}$ (3$d^{2.4}$) and Os$^{3.9+}$ (4$d^{4.1}$).
In LaSrMoTcO$_6$,
the calculated occupation numbers are 0.74 e (spin-up) and 1.82 e (spin-down)
at the Mo-site, and 2.40 e (spin-up) and 1.23 e (spin-down) on the Tc-site,
resulting in a charge configuration of Mo$^{3.4+}$ (3$d^{2.6}$) and Tc$^{3.4+}$ (4$d^{3.6}$).
In LaSrMoReO$_6$,
the calculated occupation numbers are 0.68 e (spin-up) and 1.94 e (spin-down)
at the Mo-site, and 2.14 e (spin-up) and 0.95 e (spin-down) on the Re-site,
resulting in a charge configuration of Mo$^{3.4+}$ (3$d^{2.6}$) and Re$^{3.9+}$ (5$d^{3.1}$).

The total and orbital-decomposed
DOS spectra of robust HM-AFM LaSrVOsO$_6$, LaSrMoTcO$_6$ and LaSrMoReO$_6$
are shown in Figs. 5, 6 and 7, respectively. As mentioned before, when Sr
is replaced by either Ca or Ba, the electronic structures of the 
resultant double perovskites are very similar to that of the corresponding
Sr-based double perovskite, and therefore they are not shown here.
Comparison of Fig. 5 with Fig. 3 shows that the electronic structure of LaSrVOsO$_6$
is very similar to that of LaSrVRuO$_6$ in the hypothetical HM-AFM state, with Os $5d$ playing the
role of Ru $4d$. Therefore, as for LaSrVRuO$_6$ in the AFM state,
the electronic structure of LaSrVOsO$_6$ consists of the O $2p$ dominant
lower valence band between 3.0 and 8.4 eV below $E_F$, the spin-up
Os $t_{2g}$ dominant upper valence band between 0.2 and 2.0 eV below $E_F$
and spin-down Os $t_{2g}$-V $t_{2g}$ hybridized conduction band between 2.0 eV
below $E_F$ and $\sim$1.0 eV above $E_F$ as well as the spin-up V $t_{2g}$ dominant,
Os $e_g$ dominant and V $e_g$ dominant upper conduction bands about 0.8 eV
above $E_F$ (see Fig. 5). LaSrVOsO$_6$ is half-metallic with a band gap of about 0.8 eV on
the spin-up channel. Fig. 5 shows that the lower valence band contains rather
discernable contributions from transition metal $B$ and $B'$ $d$ orbitals especially
Os 5$d$ in the lower part of the band, because of the hybridization between
O $p$ and $B$ ($B'$) $d$ orbitals. This rather strong O $p$ - $B$ ($B'$) $d$ 
hybridization pushes the V (Os) $e_g$ dominant band above the V (Os) $t_{2g}$ dominant band,
as shown in Fig. 5, resulting in a crystal field splitting of the $t_{2g}$ and
$e_g$ bands of about 2.0 eV for the VO$_6$ octahedron and of a slightly larger value for the
OsO$_6$ octahedron. This $B$ $d$ - O $p$ - $B'$ $d$ $\sigma$-bonding is the well known
superexchange coupling between the $B$ and $B'$ ions and gives rise to 
the AFM exchange interaction between the $B$ and $B'$ ions. 
However, in the conventional superexchange-type AFM insulators such as NiO, 
the majority (minority) spin $d$ orbitals of each transition metal ion are separated by
a large superexchange energy gap from
the minority (majority) spin $d$ orbitals of its nearest neighbor transition metal ions. 
In contrast, the majority spin V $t_{2g}$ and minority spin Os $t_{2g}$ orbitals 
in LaSrVOsO$_6$ are closely located in energy (Fig. 5). This gives rise to a strong
hybridization between the majority spin V $t_{2g}$ and minority spin Os $t_{2g}$ orbitals 
via O $p_{\pi}$ orbitals which results in the broad $B$ and $B'$ $t_{2g}$ hybridized
spin down conduction band in the energy range of -2.0$\sim$1.0 eV, and further strengthens
the AFM exchange coupling between the V and Os ions.
Clearly, this coupling between the V and Os $t_{2g}$ orbitals via O $p_{\pi}$ orbitals 
would allow the spin-down conduction electrons
to hop freely from a V ion to the neighboring Os ions and back to the V ion, resulting
in a lower kinetic energy. This situation is similar to the
double exchange mechanism~\cite{zen51} of the metallic ferromagnetism in the colossal magneto-resistive
manganites, and hence may be called the generalized double exchange mechanism.
The direct on-site exchange interaction would then lower the majority spin Os $t_{2g}$
band below $E_F$ and push the minority spin V $t_{2g}$ band above $E_F$, thereby
creating an insulating gap in the spin up channel (see Fig. 5).
As a result, the band gap may be called an antiferromagnetic coupling gap~\cite{jen03}.
Therefore, the antiferromagnetism in LaSrVOsO$_6$ has the dominant contributions from both the 
superexchange mechanism and the generalized double exchange mechanism (see Refs. ~\cite{sol99,sol03}
for the detailed analysis on these two mechanisms in the doped ferromagnetic manganites).
The origin of the half-metallicity
is essentially the same as that of the HM ferrimagnetic double perovskites such as
Sr$_2$FeMoO$_6$, namely, the $B$ ($t_{2g}$) - O (2$p_{\pi}$) - $B$ ($t_{2g}$) 
hybridization, as illustrated in Refs. ~\onlinecite{wu01,jen03}. It could then be infered
from this discussion that the antiferromagnetic transition temperature $T_N$ in
the HM-AFMs would be in the same order of magnitude as that 
in Sr$_2$FeMoO$_6$ ($T_c = 419$ K) \cite{kob98}, i.e., being above room temperature.

The electronic structures of LaSrMoTcO$_6$ and LaSrMoReO$_6$ are similar to that
of LaSrVOsO$_6$, as can be seen from Figs. 5-7. In particular, the electronic structure
of LaSrMoTcO$_6$ is nearly identical to that of LaSrVOsO$_6$. The discernable difference
in the electronic structure between LaSrMoTcO$_6$ and LaSrVOsO$_6$ is the
larger Mo 4$d$ contribution to the O $2p$ dominant lower valence band, indicating
a stronger Mo 4$d$ - O 2$p$ bonding in LaSrMoTcO$_6$. This may be expected because
the Mo 4$d$ orbitals are significantly more extended than that of V 3$d$. There are also 
some discernable differences in the electronic structure between 
LaSrMoReO$_6$ and LaSrVOsO$_6$. The most pronounced difference is that there is a
small dip or pseudop gap at $E_F$ in the spin-down channel. Therefore, the spin-down
band structure of LaSrMoReO$_6$ may be regarded as a semimetal with a small 
DOS of 0.6 states/eV/spin/f.u. at $E_F$. The spin-down DOS at $E_F$ furthers decreases
when Sr is replaced by Ba (see Table VII). Further differences among the electronic
strucrtures of LaSrVOsO$_6$, LaSrMoTcO$_6$ and LaSrMoReO$_6$ are minor and
they include slight differences in the energy position and bandwidth of various
bands.

\begin{table}
\caption{Calculated magnetic and electronic properties of La$A$XYO$_6$ in theoretically
determined P4/nmm structure using GGA.}
\begin{ruledtabular}
\center{(a) La$A$VOsO$_6$}
\begin{tabular}{cccccccc}
\multicolumn{2}{c}{$A = $}  & \multicolumn{2}{c}{Ca} & \multicolumn{2}{c}{Sr} & \multicolumn{2}{c}{Ba}\\ \hline
\multicolumn{2}{c}{magnetic states} & FM/NM  &  AF & FM/NM &  AF & FM/NM &  AF\\ \hline
\multicolumn{2}{c}{$m_V$ ($\mu_B$)} &  0.007 & -0.985 &  0.007 &  -1.014    &  0.0 & -1.057 \\
  \multicolumn{2}{c}{$m_{Os}$ ($\mu_B$)} & 0.070 & 0.699 &  0.060 &  0.731   &  0.0 & 0.734 \\
  \multicolumn{2}{c}{$m_t$ ($\mu_B$/f.u.)}  &  0.131   &   0.0  &  0.085 & 0.0  &   0.0 &  0.0 \\
\hline
  $N(E_F)$  & $\uparrow$ & 2.321 & 0.0  & 2.006 & 0.0  &  1.909  & 0.0 \\
  (states/eV/f.u.)& $\downarrow$ & 2.520 & 3.870    &  2.478 &  4.098 &  1.909 &  4.434 \\ \hline
  gap (eV) & $\uparrow$ &  & 0.65 &  &0.74 & & 0.82 \\
\end{tabular}

\center{(b) La$A$MoTcO$_6$}
\begin{tabular}{cccccccc}
\multicolumn{2}{c}{$A = $}  & \multicolumn{2}{c}{Ca} & \multicolumn{2}{c}{Sr} & \multicolumn{2}{c}{Ba}\\ \hline
\multicolumn{2}{c}{magnetic states} & FM/NM  &  AF & FM/NM &  AF & FM/NM &  AF\\ \hline
\multicolumn{2}{c}{$m_{Mo}$ ($\mu_B$)} &  0.0 & -1.029 &  0.0 &  -1.088    &  0.169 & -1.128 \\
  \multicolumn{2}{c}{$m_{Tc}$ ($\mu_B$)} & 0.0 &  1.125 &  0.0 &   1.176    &  0.120 &  1.213 \\
  \multicolumn{2}{c}{$m_t$ ($\mu_B$/f.u.)}  &  0.0   &   0.0  &  0.0 & 0.0  &   0.463 &  0.0 \\
\hline
  $N(E_F)$  & $\uparrow$ & 3.192 & 0.0  &  3.206 & 0.0  &  3.067  & 0.0 \\
  (states/eV/f.u.)& $\downarrow$ & 3.192 & 3.210    &  3.206 &  3.106  &  2.572 &   3.003 \\ \hline
  gap (eV) & $\uparrow$ &  & 0.52 &  &0.68  & & 0.79 \\
\end{tabular}

\center{(c) La$A$MoReO$_6$}
\begin{tabular}{cccccccc}
\multicolumn{2}{c}{$A = $}  & \multicolumn{2}{c}{Ca} & \multicolumn{2}{c}{Sr} & \multicolumn{2}{c}{Ba}\\ \hline
\multicolumn{2}{c}{magnetic states} & FM/NM  &  AF & FM/NM &  AF & FM/NM &  AF\\ \hline
\multicolumn{2}{c}{$m_{Mo}$ ($\mu_B$)} &  0.0 & -1.226 &  0.0 &  -1.264    &  0.0 & -1.301 \\
  \multicolumn{2}{c}{$m_{Re}$ ($\mu_B$)} & 0.0 &  1.168 &  0.0 &  1.199    &  0.0 & 1.230 \\
  \multicolumn{2}{c}{$m_t$ ($\mu_B$/f.u.)}  &  0.0   &   0.0  &  0.0 & 0.0  &   0.0 &  0.0 \\
\hline
  $N(E_F)$  & $\uparrow$ & 2.947 & 0.0  &  3.086 & 0.0  &  3.333  & 0.0 \\
  (states/eV/f.u.)& $\downarrow$ & 2.947 & 1.134  &  3.086 &  0.594&  3.333 & 0.175 \\ \hline
  gap (eV) & $\uparrow$ &  & 0.25 &  &0.35  & & 0.49 \\
\end{tabular}

\end{ruledtabular}
\end{table}

\begin{figure}
\includegraphics[width=8cm]{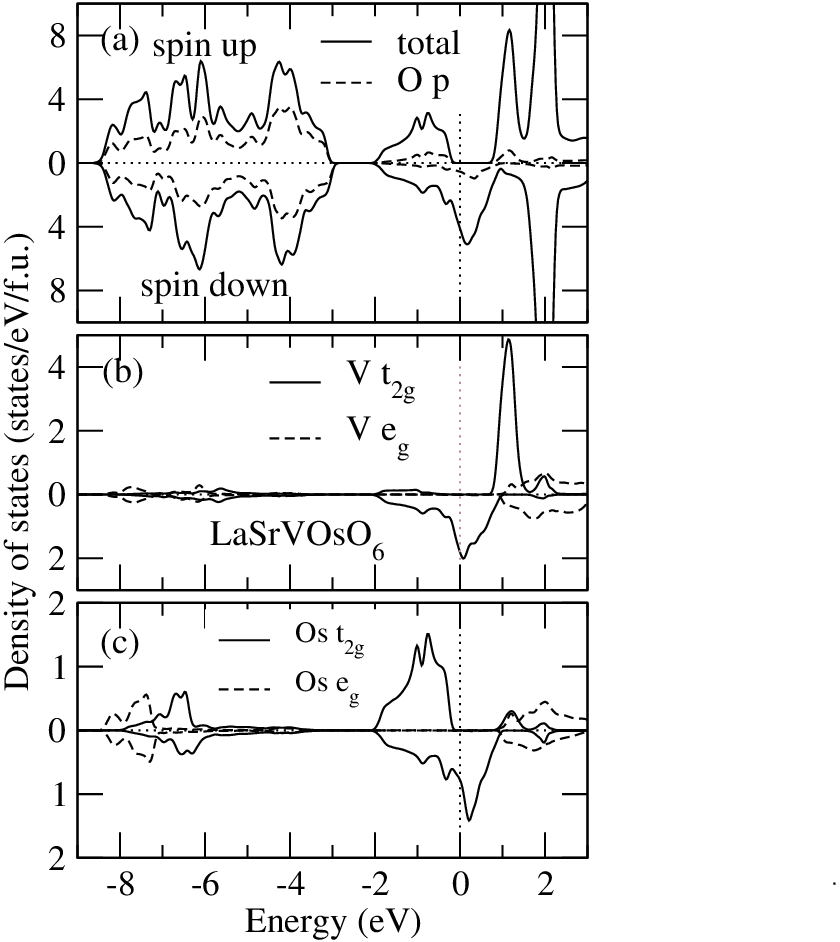}
\caption{\label{fig:fig5} Total and orbital-decomposed density of states of LaSrVOsO$_6$ 
in the theoretical determined P4/nmm structure (full structural optimization, see text).} 
\end{figure}

\begin{figure}
\includegraphics[width=8cm]{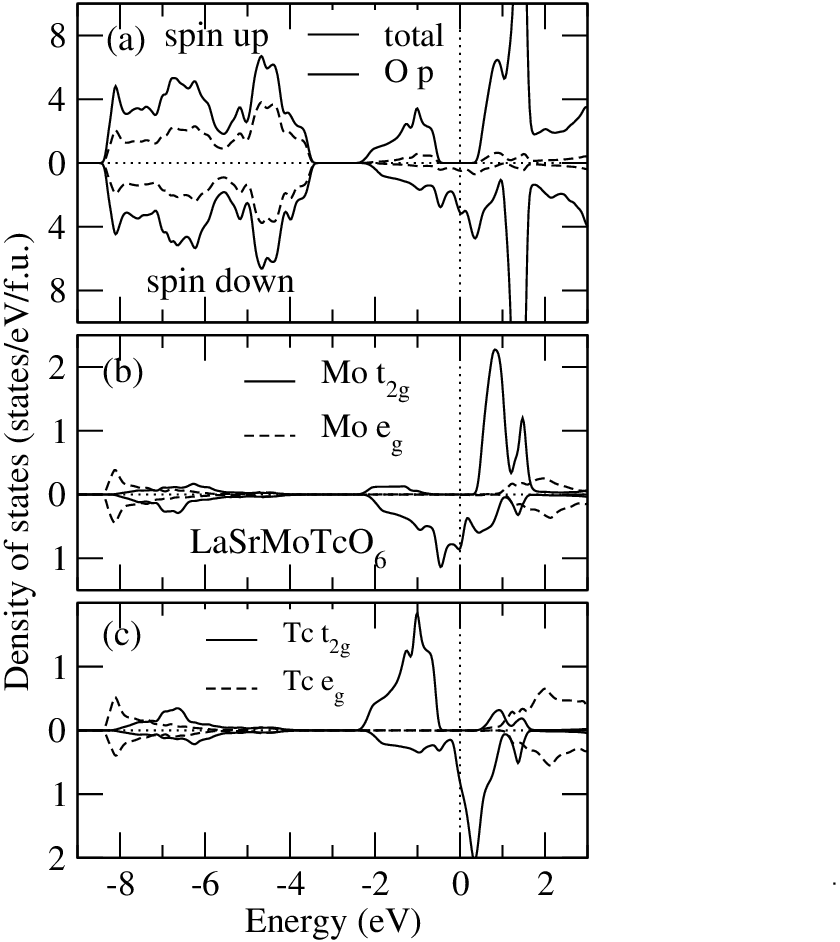} 
\caption{\label{fig:fig6} Total and orbital-decomposed density of states of LaSrMoTcO$_6$ 
in the theoretical determined P4/nmm structure (full structural optimization, see text).} 
\end{figure}

\begin{figure}
\includegraphics[width=8cm]{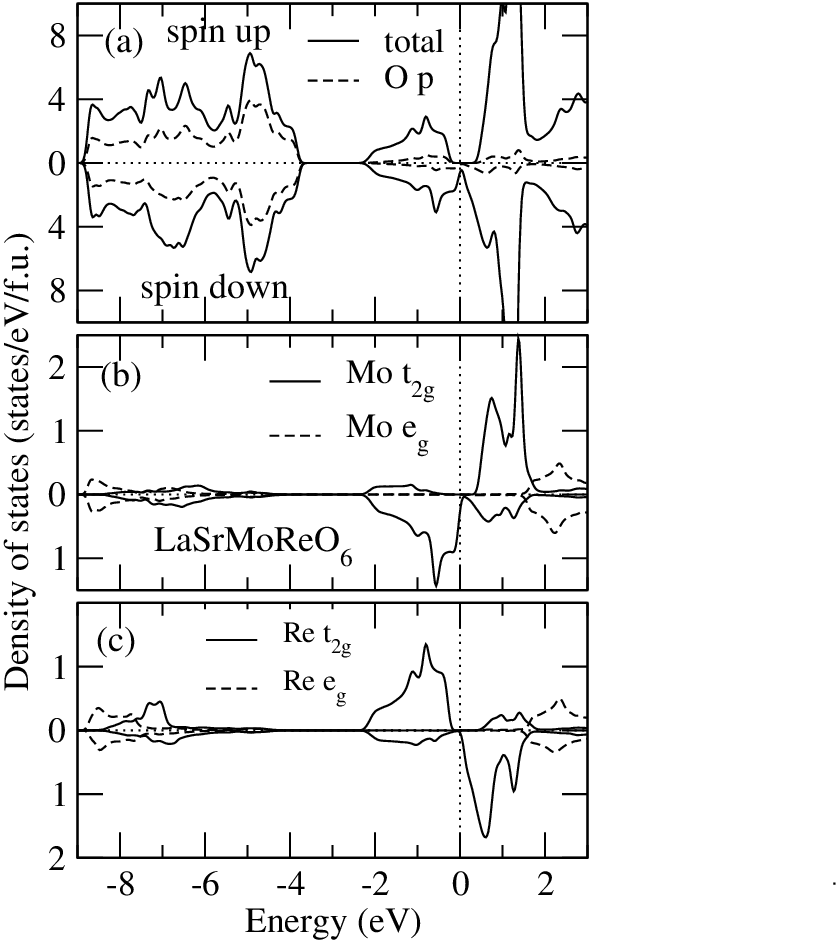} 
\caption{\label{fig:fig7} Total and orbital-decomposed density of states of LaSrMoReO$_6$ 
in the theoretical determined P4/nmm structure (full structural optimization, see text).} 
\end{figure}

\subsection{Unconventional antiferromagnetic insulators}

\begin{table}
\caption{Calculated structural parameters, magnetic and electronic properties of
antiferromagnetic insulators La$A$Cr$Y$O$_6$ in theoretically determined P4/nmm structures.}
\begin{ruledtabular}
\center{(a) La$A$CrTcO$_6$}
\begin{tabular}{ccccc}
\multicolumn{2}{c}{$A = $} &Ca & Sr & Ba\\ \hline
\multicolumn{2}{c}{$a$ (\AA)} & 5.534 & 5.573 & 5.638 \\
\multicolumn{2}{c}{$c/a$}  &1.4123 & 1.4117 & 1.4068 \\
\multicolumn{2}{c}{$V$ (\AA$^3$/f.u.)}  &119.66& 122.20 & 126.08 \\ \hline
\multicolumn{2}{c}{$m_{Cr}$ ($\mu_B$)} & -2.241 &  -2.264 &  -2.295 \\
\multicolumn{2}{c}{$m_{Tc}$ ($\mu_B$)} & 1.683 &  1.696 &  1.710 \\
\multicolumn{2}{c}{$m_t$ ($\mu_B$/f.u.)} &  0.0 & 0.0 & 0.0 \\ \hline
gap (eV) & $\uparrow$&  1.252 &  1.224  & 1.197 \\ 
   & $\downarrow$&  0.653  & 0.680  & 0.762 \\
\end{tabular}

\center{(b) La$A$CrReO$_6$} 
\begin{tabular}{ccccc}
\multicolumn{2}{c}{$A = $} &Ca & Sr & Ba\\ \hline
\multicolumn{2}{c}{$a$ (\AA)} &5.551 & 5.583 & 5.650 \\
\multicolumn{2}{c}{$c/a$}  & 1.4117 & 1.4111 & 1.4070 \\
\multicolumn{2}{c}{$V$ (\AA$^3$/f.u.)} & 120.72& 122.79 & 126.89 \\ \hline
\multicolumn{2}{c}{$m_{Cr}$ ($\mu_B$)} & -2.217 & -2.237 & -2.271\\
\multicolumn{2}{c}{$m_{Tc}$ ($\mu_B$)} &  1.468 &  1.484 & 1.503 \\
\multicolumn{2}{c}{$m_t$ ($\mu_B$/f.u.)} & 0.0 & 0.0 & 0.0 \\ \hline
gap (eV) & $\uparrow$&   0.63  & 0.76  & 0.85 \\ 
         & $\downarrow$&  0.98 &  1.09 &  1.12 \\
\end{tabular}

\end{ruledtabular}
\end{table}

In our first round of search for the HM-AFMs, we also found La$A$CrTcO$_6$ and
La$A$CrReO$_6$ be antiferromagnets. However, La$A$CrTcO$_6$ and La$A$CrReO$_6$
are insulators. Nonetheless, we may regard these materials as unconventional 
AFM insulators because they have several subtle differences in the electronic
structure from archetypical AFM insulators such as MnO and NiO in the cubic
rocksalt structure. The calculated electronic and magnetic properties of
these materials are listed in Table VIII. The total and orbital-decomposed 
DOS spectra of LaSrCrTcO$_6$ are displayed in Fig. 8, as an example.
These results are obtained from selfconsistent electronic structure calculations
for the fully theoretically optimized crystal structures. 
Compared with the HM-AFMs, these compounds have a stronger spin splitting of
the $B (B')$  $t_{2g}$ band and
a larger local magnetic moment on the $B$ and $B'$ sites 
(see Fig. 8 and Table VIII). For instance, the spin splitting of the Cr $t_{2g}$
band in LaSrCrTcO$_6$ is about 2.5 eV, significantly larger than that of the
V $t_{2g}$ band of less than 1.5 eV in LaSrVOsO$_6$. The local magnetic moment
on the Cr site in La$A$CrTc(Re)O$_6$ is twice as large as that on the V site
in La$A$VOsO$_6$. Furthermore, the spin-down Cr $t_{2g}$ band in 
La$A$CrTc(Re)O$_6$ is aligned with the spin-up Tc(Re) $t_{2g}$ band, rather
than with the spin-down Tc(Re) $t_{2g}$ band as is the case in the HM-AFM 
compounds discussed in the previous Subsec. Consequently, in La$A$Tc(Re)O$_6$,
the generalized double exchange mechanism is ineffective and the antiferromagnetic 
coupling is due solely to the traditional superexchange mechanism.
This gives rise to the AFM insulating behavior in these compounds.

\begin{figure}
\includegraphics[width=8cm]{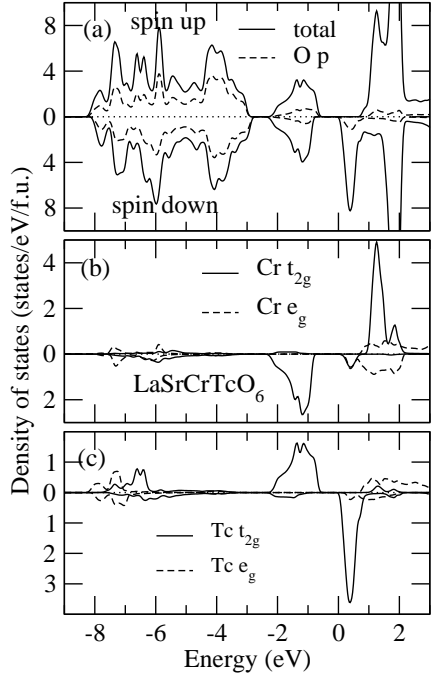}
\caption{\label{fig:fig8} Total and orbital-decomposed density of states of LaSrCrTcO$_6$ in 
the theoretical determined P4/nmm structure.} \end{figure}

In the conventional AFM insulators, the two antiferromagnetic coupled ions
are made of the same atomic species, e.g., Ni in NiO. The two spin-decomposed
total DOS spectra are identical and the insulating gaps for spin up and spin down
channels are the same. 
The novel features of the present AFM insulators include
that the two spin-decomposed total DOS spectra are no longer the same and 
the spin up insulating gap may have a size different from that of the
spin down insulating gap, as shown in Table VIII and Fig. 8, and that
the antiferromagnetic coupled ions consist of two different atomic species,
e.g., Cr and Tc in La$A$CrTcO$_6$.

\section{CONCLUSIONS}
In conclusion, in order to search for the fascinating half-metallic
antiferromagnetic materials, we have carried out a systematic {\it ab initio} 
study of the ordered double perovskites La$ABB'$O$_6$ with the 
possible $B$ and $B'$ pairs from all the 3$d$, 4$d$ and 5$d$ transtion metal 
elements being considered.
Electronic structure calculations based on first-principles density-functional
theory with GGA for more than sixty double
perovskites LaCa$BB'$O$_6$ have been performed using the all-electron
FLAPW method.
As a result, we find three families of the HM-AFMs, 
namely, La$A$VOsO$_6$, La$A$MoTcO$_6$ and La$A$MoReO$_6$ ($A$ = Ca, Sr, Ba). 
The found HM-AFM state in these materials subsequently survives the full {\it ab initio}
lattice constant and atomic position optimizations which were carried out
using frozen-core full potential PAW method.
We attribute the AFM to both the superexchange mechanism and the generalized double
exchange mechanism via the $B$ ($t_{2g}$) - O (2$p_{\pi}$) - $B'$ ($t_{2g}$)
coupling and also believe the latter to be the origin of the HM.
We also find that the HM-AFM properties predicted 
previously in some of the double perovskites would disappear after the 
full structural optimizations,
thereby providing an explanation why the previous experimental attempts to
synthesize the HM-AFM double perovskites did not succeed. 
Finally, in our search for the HM-AFMs,
we find La$A$CrTcO$_6$ and La$A$CrReO$_6$ to be AFM insulators of an unconventional
type in the sense that the two antiferromagnetic coupled ions consist of two
different elements and that the two spin-resolved densities of states are no longer
the same. Stimulated by previous theoretical predictions of
the HM-AFM in La$_2$MnVO$_6$\cite{pic98} and also in La$A$VRuO$_6$\cite{par02},
experimental effort to synthesize the HM-AFMs has been reported 
by Androulakis, {\it et al,}\cite{and02}, and attempted by Liu, 
{\it et al,}\cite{liu02}, respectively, which were, however, unsuccessful.
It is hoped that our predictions of robust HM-AFMs would encourage further
experimental searches for the HM-AFMs.

\section*{ACKNOWLEDGEMENTS}
The authors thank Ru Shi Liu for helpful discussions on the possibility of 
synthesis of the half-metallic antiferromagnet double perovskites.
The authors gratefully acknowledge financial supports from National Science Council
of Taiwan, and NCTS/TPE. They also thank National Center for High-performance
Computing of Taiwan for providing CPU time.\\

\end{document}